\documentstyle[pra,epsf,aps,twocolumn]{revtex}
\begin{document}
\draft
\title{Comment on "Continuum dual theory of the transition in 3D lattice
superconductor"}
\author{Hagen Kleinert and Adriaan M.J. Schakel}
\address{Institut f\"ur Theoretische Physik \\
Freie Universit\"at Berlin \\
Arnimallee 14,
114195 Berlin,
Germany
\thanks{
Email:
kleinert@physik.fu-berlin.de,
schakel@physik.fu-berlin.de;
URL: http://www.physik.fu-berlin.de/\~{}kleinert. Phone/Fax: 00-49-30-838 3034
}
}
\date{\today}
\maketitle
\begin{abstract}
A recent paper by Herbut in J. Phys. A: Math. Gen. {\bf 29}, 1
(1996) is shown to contain
an internal inconsistency which invalidates the principal
conclusion of the paper that the magnetic penetration depth diverges with an
$XY$-exponent rather than a mean-field exponent, as
predicted some time ago by Kiometzis,
Kleinert, and Schakel (KKS).
\end{abstract}

In a recent paper, Herbut \cite{Herbut} studies the critical behavior of the
three-dimensional superconducting phase transition using a
disorder field theory dual to
the Ginzburg-Landau theory. The
dual theory
describes a superconductor via
a disorder field $\psi$ rather than an order field \cite{GFCM}.  The field
$\psi$ accounts for the fluctuating magnetic vortex loops in the
superconducting
phase, and acquires a nonzero expectation value {\em above\/} the critical
temperature $T_c$ indicating the condensation of these loops.
The dual theory
also
contains a vector field which is massive below $T_c$ and may be identified with
the screened magnetic field.
For a lattice superconductor,
a dual formulation
was found in
\cite{PTS}. The disorder field theory
was derived in \cite{lett},
and served to
predict
the existence of a tricritical point in the superconductive phase transition,
a prediction which was confirmed by Monte
Carlo simulations in \cite{M}.  The disorder hteory was further investigated in
\cite{KKR}, and derived in an alternative
way from the Ginzburg-Landau theory in \cite{MA}.
In mean-field approximation, the
vector field decouples at the second-order
superconductive-normal transition \cite{lett}, and this is true also
in the presence of fluctuations, as shown in
a renormalization group analysis of
Kiometzis, Kleinert, and Schakel (KKS) in Ref.~\cite{KKS}.  At the critical
point, the disorder field theory becomes a pure $\psi^4$-theory
whose critical exponents are of the $XY$ universality class
(on a reversed temperature axis).
  In particular,
exponent $\nu$
in the temperature behavior $|T-T_c|^{- \nu}$
of the coherence length
of the disorder field is $\nu \approx 2/3$,
implying a specific heat behavior $|T-T_c|^{- \alpha}$
with $ \alpha=2-3  \nu\approx 0$.

Since the magnetic vector field decouples at the critical point, it does not
participate intensively enough in the critical fluctuations to acquire
nontrivial critical exponents.  For this reason, the magnetic penetration depth
$\lambda(T)$ diverges near $T_c$ with a {\em mean-field} exponent.  This
surprising prediction has apparently
been confirmed in experiments by Lin {\it et al.}
\cite{Lin} on YBaCuO.  The confirmation is not yet
beyond doubts, however, since it relies on a delicate
finite-size analysis,
leaving room
for a reinterpretation of the
data and the possibility of
suggesting a different critical behavior.

This was done by Herbut \cite{Herbut} who criticized the
renormalization group
treatment of \cite{KKS}, arguing that the magnetic penetration depth
$\lambda(T)$ should have the same $XY$-exponent as the coherence length
$\xi(T)$
of the disorder field.  The purpose of this Comment is to point out
inconsistencies in his work which invalidate his conclusion.

Below Eq.~(16) in Ref.~\cite{Herbut},
Herbut shows that upon approaching the
critical point, the mass $m_h$ of the massive vector field
becomes infinitely
heavy as compared to the mass $m$ of the disorder field.  More specifically ,
he
shows that $m^2_h/m \rightarrow {\rm const.}$, implying that $m_h/m \rightarrow
\infty$.
Thus the vector field decouples from the critical theory, in agreement with
KKS.  In the subsequent paragraph, however,
the author
argues that $\lambda$ diverges in the same way as the disorder correlation
length
$\xi = m^{-1}$, implying that the ratio $\lambda^{-1}/m$
approaches a {\em finite\/} limit.
This means that $m_h$ cannot be identified
with the inverse penetration length
as one would expect
because of the physical meaning of the massive vector
field.
The length $m_h^{-1}$ would
constitute a third length scale in the dual
formulation, although physically only two are present.
The conclusion that
$\lambda \propto \xi$ is arrived at by incorporating renormalization group
results of the Ginzburg-Landau theory.  But this is
 not permitted.  The dual
formulation constitutes an independent description of the phase transition,
and
all renormalization group results should be derived within this framework.

A second inconsistency in Herbut's paper
 is that he argues that the critical
point is characterized by a nonzero dual charge rather than a
vanishing one as
is the case in KKS.  This is in disagreement with the global phase diagram
derived in Ref.~\cite{KKR} on general grounds.  In fact, Herbut's conclusion
would imply the presence of two fixed points, one with zero and one with
nonzero
dual charge, both characterized by the same $XY$ disorder exponents.

The two inconsistencies can be traced back to Herbut's faulty treatment of the
penetration length at the mean-field level.
It is known experimentally that
outside the critical region, $\lambda$ has the temperature dependence $(T -
T_{\rm c})^{-1/2}$.  This has to be properly accounted for by the parameters of
the bare disorder field theory, as was done in KKS
Any
renormalization group calculation of fluctuation corrections
must start from the observable mean-field behavior
in the wider
neighborhood of $T_c$.
The renormalization group is
only capable of producing {\em corrections\/}
to the mean-field behavior, not that behavior itself \cite{rem}.
These corrections make
 $m_h/m$ approach infinity at
the critical point rather than a constant,
as in the mean-field approximation.
They do not influence
the
vanishing of
the dual charge in
the
mean-field theory,
as shown by KKS.
With the identification
$\lambda = m_h^{-1}$,
the dual formulation contains
only two length scales---as is physically the case.

The simple physical picture emerging from this theory is that the critical
theory is a pure disorder $\psi^4$-theory with $XY$-exponents.  A
massive vector field accompanies the transition
as a spectator, without changing the
critical exponents.
Although the vector field becomes itself
massless at the
critical point, its fluctuations decouple
at
the phase transition,
and
its penetration depth diverges with a mean-field like
exponent.

In closing, we remark that we disagree with Herbut's statement that Eq.~(15) of
his work is in accordance with the Josephson relation.  That relation links the
Cooper-pair density to the correlation length of the {\it order} field
describing the Cooper-pair condensate, not the {\it disorder} field describing
vortices which appears in his Eq.\ (15).

\end{document}